\documentclass[aps,prb,floatfix,twocolumn,showpacs]{revtex4}%
\usepackage{amsmath}
\usepackage{graphicx}
\usepackage{amsfonts}
\usepackage{amssymb}%
\providecommand{\U}[1]{\protect\rule{.1in}{.1in}}
\providecommand{\U}[1]{\protect\rule{.1in}{.1in}}
\providecommand{\U}[1]{\protect\rule{.1in}{.1in}}
\providecommand{\U}[1]{\protect\rule{.1in}{.1in}}
\providecommand{\U}[1]{\protect\rule{.1in}{.1in}}

\begin{document}
\title{Probing coherent Cooper pair splitting with cavity photons}
\author{Audrey Cottet}
\affiliation{Laboratoire Pierre Aigrain, Ecole Normale Sup\'{e}rieure, CNRS UMR 8551,
Laboratoire associ\'{e} aux universit\'{e}s Pierre et Marie Curie et Denis
Diderot, 24, rue Lhomond, 75231 Paris Cedex 05, France}

\begin{abstract}
This work discusses theoretically the behavior of a microwave cavity and a
Cooper pair beam splitter (CPS) coupled non-resonantly. The cavity frequency
pull is modified when the CPS is resonant with a microwave excitation. This
provides a direct way to probe the coherence of the Cooper pair splitting
process. More precisely, the cavity frequency pull displays an anticrossing
whose specificities can be attributed unambiguously to coherent Cooper pair
injection. This work illustrates that microwave cavities represent a powerful
tool to investigate current transport in complex nanocircuits.

\end{abstract}

\maketitle

\section{Introduction}

Superconductors represent a natural source of entanglement due to Cooper pairs
which gather two electrons in the spin singlet state. The spatial separation
of these electrons is an interesting goal in the context of quantum
computation and communication. In principle, a Cooper pair beam splitter (CPS)
connected to a central superconducting contact and two outer normal metal (N)
contacts could facilitate this process\cite{Recher:01}. The spatial splitting
of Cooper pairs has been demonstrated experimentally from an analysis of the
CPS average currents, current noise and current
cross-correlations\cite{CPSexp,Schindele}. However, new tools appear to be
necessary to investigate further the CPS dynamics, and in particular its
coherence, which has not been demonstrated experimentally so
far\cite{NewTools,Schroer}. This coherence has two intimately related aspects:
the coherence of Cooper pair injection and the conservation of
spin-entanglement. The first aspect is due to the fact that Cooper pair
injection into the CPS is a coherent crossed Andreev process, which produces a
coherent coupling between the initial and final states of the Cooper pair in
the superconducting contact and the CPS (see e.g. \cite{Sauret,Eldridge}). The
observation of coherent pair injection appears as an important prerequisite
for the realization of a fully coherent CPS.

In Cavity Quantum Electrodynamics (QED)\cite{QED,Raimond} or Circuit
QED\cite{Wallraff}, real or artificial two levels atoms are controlled and
readout with a high accuracy thanks to the use of cavity photons. Very
recently, coplanar microwave cavities have been coupled to nanocircuits based
on carbon nanotubes (CNTs), semiconducting nanowires or two-dimensional
electron gases\cite{Delbecq,DQD,Petersson,Viennot}. This paves the way for the
development of a Hybrid Circuit QED which offers many possibilities due to the
versatility of nanocircuits made with nanolithography techniques. Indeed,
nanoconductors can be coupled to various types of reservoirs such as normal
metals, ferromagnets\cite{CottetSST} or superconductors\cite{Franceschi}, in a
large variety of
geometries\cite{Qubits,CKLY,MajoDirect,MajoTransmon,Lasing,2DQDS}. Hybrid
Circuit QED tackles problems which go beyond the mechanics of closed two level
systems. In particular, the interaction between electronic transport and the
light-matter interaction leads to a rich
phenomenology\cite{CKLY,Lasing,2DQDS,Liu}. Photon emission in the
cavity/nanoconductor resonant regime has received the most attention so far.
In contrast, this work considers a CPS and a cavity coupled non-resonantly, so
that the CPS simply causes a cavity frequency pull. When the CPS is excited
with a microwave voltage, the cavity frequency pull displays an anticrossing
which can be attributed unambigously to coherent Cooper pair injection, due to
various specificities related to the transport geometry and the symmetries of
the split singlet Cooper pairs. More generally, this work illustrates that
Hybrid Circuit QED provides a powerful tool to investigate current transport
in complex nanocircuits.

\section{Hamiltonian description of the CPS and cavity}

I consider a CNT (light blue) placed between the center and ground conductors
(purple) of a superconducting coplanar waveguide cavity (Fig. 1.a). A grounded
superconducting contact (purple) and two outer N contacts (black) biased with
a voltage $V_{b}$ are used to define two quantum dots $L$ and $R$ along the
CNT. The dot $L(R)$ is placed close to a gate electrode (gray) biased with a
DC voltage $V_{g}^{L(R)}$. I use the CPS hamiltonian%
\begin{align}
H_{CPS} &  =%
%TCIMACRO{\dsum \nolimits_{i,\tau,\sigma}}%
%BeginExpansion
{\displaystyle\sum\nolimits_{i,\tau,\sigma}}
%EndExpansion
\left(  (\varepsilon+\Delta_{so}\tau\sigma)d_{i\tau\sigma}^{\dag}%
d_{i\tau\sigma}+\frac{\varepsilon_{B}}{2}d_{i\tau\sigma}^{\dag}d_{i\tau
\overline{\sigma}}\right)  \nonumber\\
&  +\Delta_{K\leftrightarrow K^{\prime}}%
%TCIMACRO{\dsum \nolimits_{i,\sigma}}%
%BeginExpansion
{\displaystyle\sum\nolimits_{i,\sigma}}
%EndExpansion
(d_{iK\sigma}^{\dag}d_{iK^{\prime}\sigma}+h.c.)\label{H}\\
&  +t_{ee}%
%TCIMACRO{\dsum \nolimits_{\tau,\sigma}}%
%BeginExpansion
{\displaystyle\sum\nolimits_{\tau,\sigma}}
%EndExpansion
(d_{L\tau\sigma}^{\dag}d_{R\tau\sigma}+h.c.)+H_{prox}+H_{int}\nonumber
\end{align}
with
\begin{equation}
H_{prox}=t_{eh}%
%TCIMACRO{\dsum \nolimits_{\tau}}%
%BeginExpansion
{\displaystyle\sum\nolimits_{\tau}}
%EndExpansion
\left(  d_{L\tau\uparrow}^{\dag}d_{R\overline{\tau}\downarrow}^{\dag
}-d_{L\overline{\tau}\downarrow}^{\dag}d_{R\tau\uparrow}^{\dag}\right)
+h.c.\label{Hprox}%
\end{equation}
The operator $d_{i\tau\sigma}^{\dag}$ creates an electron with spin $\sigma
\in\{\uparrow,\downarrow\}$ along the CNT axis, in orbital $\tau
\in\{K,K^{\prime}\}$ of dot $i\in\{L,R\}$. The twofold orbital degeneracy is
due to the atomic structure of the CNT. The term in $\Delta_{so}$ is caused by
spin-orbit coupling\cite{Jespersen}. The term in $\Delta_{K\leftrightarrow
K^{\prime}}$ describes a coupling between the $K$ and $K^{\prime}$ orbitals,
due to disorder in the CNT atomic structure
\cite{Liang,Kuemmeth,Jespersen,Palyi}. The term in $t_{ee}$ describes interdot
hopping. An external magnetic field $\overrightarrow{B}$ is applied in the
plane of the cavity, perpendicular to the CNT. This produces a Zeeman
splitting $\varepsilon_{B}=g\mu_{B}B$ of the spin states in the dots. The term
$H_{int}$ describes Coulomb interactions inside the CPS. In this work, it is
assumed that the local Coulomb charging energy in each dot is very large so
that a dot cannot be doubly occupied. The term $H_{prox}$ accounts for the
proximity effect caused by the superconducting contact. More precisely, it
describes the coherent injection of singlet Cooper pairs inside the CPS, due
to non-local Andreev reflections. Note that in principle, $H_{prox}$ should
also include terms $\Delta_{loc,i}\left(  d_{i\tau\uparrow}^{\dag
}d_{i\overline{\tau}\downarrow}^{\dag}-d_{i\overline{\tau}\downarrow}^{\dag
}d_{i\tau\uparrow}^{\dag}\right)  $, with $i\in\{L,R\}$, describing intra-dot
pairing and local Andreev reflections. However, these terms are not relevant
in this work due to the assumption of large intra-dot Coulomb interaction. The
use of $H_{prox}$ instead of a full microscopic description of the
superconducting contact also requires one to consider subgap bias voltages,
for which single quasiparticle transport between the superconducting contact
and the dots is forbidden\cite{Eldridge}.

The total hamiltonian describing the CPS and the cavity is%
\begin{equation}
H_{tot}=H_{CPS}+\hbar\omega_{cav}a^{\dag}a+H_{c}+H_{bath}%
\end{equation}
where $a^{\dag}$ creates a cavity photon. \begin{figure}[ptb]
\includegraphics[width=1.\linewidth]{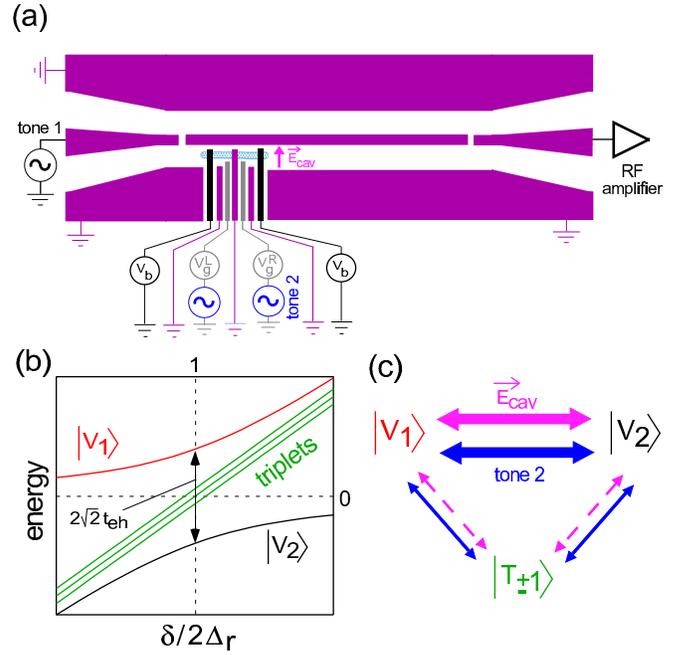}\newline\caption{(a) Scheme of
the CPS embedded in a\ coplanar microwave cavity (for details, see text). (b)
Energy levels of the subspace $\mathcal{E}$ near $\delta\sim2\Delta_{r}$ (c)
Scheme of the couplings between some states of $\mathcal{E}$, provided by the
cavity electric field (pink arrows) and the classical tone 2 (blue arrows).
The couplings corresponding to the dashed arrows can be disregarded in this
study (see text).}%
\end{figure}The term $H_{c}$ describes the CPS/cavity coupling and $H_{bath}$
describes the coupling of the CPS and cavity to dissipative baths, including
the N reservoirs and their DC voltage bias. The amplitude of the cavity
electric field can be expressed as $E_{cav}=V_{rms}(a+a^{\dag})/\ell$ with
$V_{rms}$ a characteristic voltage and $\ell$ the distance between the ground
and center conductors of the cavity. Due to the imperfect screening of
$E_{cav}$ by the CNT, the coupling between the CPS and the cavity can occur
through three paths, i.e. $H_{c}=h_{c}(a+a^{\dag})$ with%
\begin{equation}
h_{c}=%
%TCIMACRO{\dsum \limits_{i,\tau,\sigma}}%
%BeginExpansion
{\displaystyle\sum\limits_{i,\tau,\sigma}}
%EndExpansion
\left(  \beta_{i}n_{i\tau\sigma}d_{i\tau\sigma}^{\dag}d_{i\tau\sigma
}+\mathbf{i}\sigma\lambda_{i}d_{i\tau\sigma}^{\dag}d_{i\tau\overline{\sigma}%
}+\alpha_{i}d_{i\tau\sigma}^{\dag}d_{i\bar{\tau}\sigma}\right)  \label{coupl}%
\end{equation}
The first term of $H_{c}$ describes a shift of the chemical potential of dot
$i$ proportionally to the cavity electric field $\vec{E}_{cav}$. The
second[third] term describes a coupling of the electrons motion to $\vec
{E}_{cav}$, which enables photon-induced spin-flips [orbit-changes] due to
spin-orbit interaction [atomic disorder] in the CNT\cite{CKLY,Palyi}. The
coefficients $\beta_{i}$, $\lambda_{i}$ and $\alpha_{i}$ can be calculated
microscopically in a consistent way, by assuming for instance that $\vec
{E}_{cav}$ is uniform on the scale of the CPS \cite{CKD}.

Due to $H_{int}$, it is possible to tune $V_{g}^{L(R)}$ such that there is a
single electron on each dot when $V_{b}=0$ and $t_{eh}=0$. I denote with
$\delta$ the charging energy corresponding to such an occupation, with respect
to the charging energy for having the CPS empty state $\left\vert
0,0\right\rangle $. One can tune $\delta$ with $V_{g}^{L(R)}$. When $t_{eh}$,
$\varepsilon_{B}\ll\Delta_{r}$ and $\delta\sim2\Delta_{r}$ with $\Delta
_{r}=\sqrt{\Delta_{so}^{2}+\Delta_{K\leftrightarrow K^{\prime}}^{2}}$, one can
isolate an ensemble $\mathcal{E}=\{\left\vert V_{1}\right\rangle ,\left\vert
V_{2}\right\rangle ,\left\vert T_{+}\right\rangle ,\left\vert T_{-}%
\right\rangle ,\left\vert T_{0}\right\rangle \}$ of five CPS even-charged
eigenstates which are below all other even-charged eigenstates, by an energy
$\sim2\Delta_{r}$\ at least. The eigenstates $\left\vert V_{1}\right\rangle $
and $\left\vert V_{2}\right\rangle $ are a coherent superposition of
$\left\vert 0,0\right\rangle $ and a spin singlet state $\left\vert
\mathcal{S}\right\rangle $, due to the term in $t_{eh}$. The states
$\left\vert \mathcal{S}\right\rangle $ and $\left\vert T_{n}\right\rangle $,
with $n\in\{-1,0,1\}$, are generalized spin singlet and spin triplet states,
whose definition takes into account the existence of the $K/K^{\prime}$
orbital degeneracy (see Appendix A). The energy of the different states of
$\mathcal{E}$ is given by%
\begin{equation}
E_{V_{1(2)}}=\frac{1}{2}\left(  \delta-2\Delta_{r}\pm\sqrt{8t_{eh}^{2}%
+(\delta-2\Delta_{r})^{2}}\right)  \label{11}%
\end{equation}
and%
\begin{equation}
E_{T_{n}}=\delta-2\Delta_{r}+n\frac{\Delta_{K\leftrightarrow K^{\prime}}%
}{\Delta_{r}}\varepsilon_{B}\label{Et}%
\end{equation}
As visible in Eq.(\ref{11}), the states $\left\vert V_{1}\right\rangle $ and
$\left\vert V_{2}\right\rangle $ form an anticrossing with a width $2\sqrt
{2}t_{eh}$ at $\delta\sim2\Delta_{r}$ (see Fig.1.b). This anticrossing
directly reveals the coherence of the Cooper pair injection process. It is
thus crucial to be able to identify this feature in an experiment. In this
work, we show that the microwave cavity represents a powerfull tool to perform
this task.

The states of $\mathcal{E}$ are coupled by cavity photons. I denote with
$\sigma_{cd}$ the transition operator from states $d$ to $c$ and $\omega
_{cd}=(E_{c}-E_{d})/\hbar$. Inside $\mathcal{E}$, the cavity/CPS coupling is
written as
\begin{equation}
H_{c}=eV_{rms}%
%TCIMACRO{\tsum \nolimits_{cd}}%
%BeginExpansion
{\textstyle\sum\nolimits_{cd}}
%EndExpansion
\alpha_{cd}\sigma_{cd}(a+a^{\dag})
\end{equation}
with \begin{figure}[ptb]
\includegraphics[width=1.\linewidth]{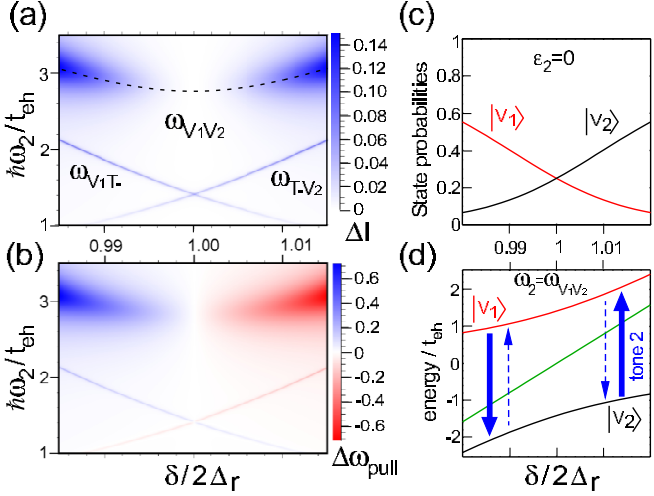}\newline\caption{(a) Current
variation $\Delta I$ versus $\delta$ and $\omega_{2}$ for $\varepsilon_{B}=0$
and a negligible relaxation between the states of $\mathcal{E}$ (b)
Corresponding $\Delta\omega_{pull}$ (c) Occupation probabilities of states
$\left\vert V_{1}\right\rangle $ and $\left\vert V_{2}\right\rangle $ for
$\varepsilon_{2}=0$ (d) Scheme illustrating that tone 2 lifts the current
blockade through the CPS at both sides of $\delta=2\Delta_{r}$. We have used
the realistic parameters $t_{eh}=12~\mathrm{\mu eV}$, $\Delta_{so}%
=0.15~\mathrm{meV}$, $\Delta_{K/K^{\prime}}=0.45~\mathrm{meV}$, $\Gamma
_{N}=125~$\textrm{M}$\mathrm{Hz}$, $\omega_{cav}=2\pi\times10~$\textrm{G}%
$\mathrm{Hz}$, $V_{rms}=4~\mathrm{\mu eV}$, $\varepsilon_{2}=150~\mathrm{\mu
eV}$, $\beta_{L(R)}=10^{-2}$, $\lambda_{L}-\lambda_{R}=10^{-4}$, and
$\alpha_{L(R)}\ll\beta_{L(R)}$. In all the Figs. of this paper, $\Delta I$ is
reduced by $e\Gamma_{N}$ and $\Delta\omega_{pull}$ is reduced by the scale
$\omega_{0}$ defined in Eq. (\ref{om0}).}%
\end{figure}%
\begin{equation}
\alpha_{T_{\pm}V_{1(2)}}=\mp v_{1(2)}\mathbf{i}(\lambda_{L}-\lambda_{R}%
)\Delta_{K\leftrightarrow K^{\prime}}/\Delta_{r}\sqrt{2}\label{alpha1}%
\end{equation}%
\begin{equation}
\alpha_{T_{\pm}T_{_{0}}}=\mathbf{i}(\lambda_{L}+\lambda_{R})\Delta
_{K\leftrightarrow K^{\prime}}/\Delta_{r}\sqrt{2}\label{alpha1b}%
\end{equation}%
\begin{equation}
\alpha_{V_{_{2}}V_{_{1}}}=v_{1}v_{2}\left[  (\beta_{L}+\beta_{R})-\left(
(\alpha_{L}+\alpha_{R})\Delta_{K\leftrightarrow K^{\prime}}/\Delta_{r}\right)
\right]  \label{alpha2}%
\end{equation}
$\alpha_{T_{0}V_{1(2)}}=0$ and $\alpha_{cd}=\alpha_{cd}^{\ast}$. The term
(\ref{alpha1}) displays destructive interferences between the spin-flip
coupling elements $\lambda_{L}$ and $\lambda_{R}$ because it describes
transitions between singlet and triplet states\cite{CKLY}. In contrast,
(\ref{alpha1b}) depends on $\lambda_{L}+\lambda_{R}$ because it describes
transitions between triplet states. The term (\ref{alpha2}) depends on
$\beta_{L}+\beta_{R}$ because it involves transitions between $\left\vert
0,0\right\rangle $ and $\left\vert \mathcal{S}\right\rangle $, which are
triggered by a common oscillation of the two dot levels with respect to the
potential of the superconducting contact. It also displays a constructive
interference between $\alpha_{L}$ and $\alpha_{R}$.

\section{Cavity frequency pull and CPS input current in the CPS/cavity
non-resonant regime}

\subsection{Description of the measurement scheme}

Since the couplings $\lambda_{L(R)}$ are expected to be weak, the effects of
(\ref{alpha1}) and (\ref{alpha1b}) should be measurable only when the cavity
is closely resonant with the CPS. For instance, Ref.\cite{CKLY} discusses a
lasing effect which occurs when $\omega_{V_{1}T_{\mp}}=\omega_{cav}$. Such an
effect could be challenging to observe because it requires reaching a lasing
threshold. For that purpose, it could be necessary to use a cavity with a high
quality factor $Q\geq10^{6}$, not achieved yet in Hybrid Circuit QED. This
work discusses the opposite regime, i.e. the cavity and the CPS are
non-resonant, so that the CPS can only produce a cavity frequency pull
$\omega_{pull}$. This effect is due to an exchange of virtual photons between
the CPS and the cavity. Since $\beta_{L(R)}\gg\lambda_{L[R]}$, $\alpha_{L[R]}$
is expected, one can neglect the contribution of (\ref{alpha1}) and
(\ref{alpha1b}) to $\omega_{pull}$. At second order in $\alpha_{V_{_{2}%
}V_{_{1}}}$, one finds%

\begin{equation}
\omega_{pull}=\mathcal{C}\omega_{0}(P_{V_{1}}-P_{V_{2}})\label{pull}%
\end{equation}
where the probability $P_{V_{1(2)}}$ of state $\left\vert V_{1(2)}%
\right\rangle $ can be calculated in the absence of the cavity, and the
parameters $\mathcal{C}$ and $\omega_{0}$ are defined as
\begin{equation}
\mathcal{C}=-\frac{2\omega_{cav}\omega_{V_{1}V_{2}}}{\omega_{cav}^{2}%
-\omega_{V_{2}V_{1}}^{2}}%
\end{equation}
and
\begin{equation}
\omega_{0}=(\alpha_{V_{_{2}}V_{_{1}}}eV_{rms})^{2}/\omega_{cav}\label{om0}%
\end{equation}
In practice, $\omega_{pull}$ can be obtained by measuring the cavity response
to a weak microwave drive (tone 1) $H_{d,1}=\varepsilon_{1}e^{i\omega_{1}%
t}a+h.c.$ with frequency $\omega_{1}\sim\omega_{cav}$\cite{Wallraff}. This
will not modify $P_{V_{1(2)}}$ since the cavity and the CPS states are off
resonant. Meanwhile, a second microwave drive (tone 2) $H_{d,2}=%
%TCIMACRO{\tsum \nolimits_{cd}}%
%BeginExpansion
{\textstyle\sum\nolimits_{cd}}
%EndExpansion
\varepsilon_{2,cd}e^{i\omega_{2}t}\sigma_{cd}+h.c.$ with frequency $\omega
_{2}$ can be applied on the CPS gates to control directly the CPS state. For
simplicity, one can assume that the electric field $\vec{E}_{2}$ associated
with tone 2 is parallel to $\vec{E}_{cav}$ and uniform on the scale of the
CPS, so that one can use $\varepsilon_{2,cd}=\varepsilon_{2}\alpha_{cd}$, with
$\varepsilon_{2}=e\ell E_{2}$. One cannot disregard the elements
$\varepsilon_{2,cd}$ involving $\left\vert T_{+}\right\rangle $ or $\left\vert
T_{-}\right\rangle $ because $\omega_{2}$ can be resonant with any of the CPS
transitions.\begin{figure}[ptb]
\includegraphics[width=1.\linewidth]{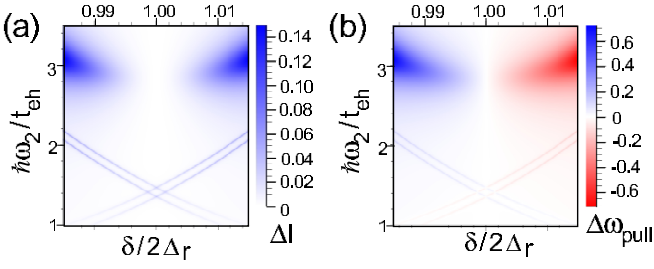}\newline\caption{(a) Current
variation $\Delta I$ versus $\delta$ and $\omega_{2}$ for $B$ finite. (b)
Corresponding $\Delta\omega_{pull}$. We have used the parameters of Fig.2 and
$\varepsilon_{B}=0.7~\mathrm{\mu eV}$.}%
\end{figure}

The present work describes how tone 2 modifies $\omega_{pull}$ and the average
current $I_{CPS}$ flowing through the CPS superconducting contact for $V_{b}$
finite. I consider a range of $V_{b}$ and $\delta$ such that electrons can go
from the dots to the N reservoirs but not the reverse, and transport processes
involve only the states from $\mathcal{E}$ and the CPS singly occupied
states\cite{CKLY}. Assuming that the bare coupling rate $\Gamma_{N}$ between
the dots and the N reservoirs is independent from $i$, $\tau$, and $\sigma$,
the details on the CPS singly occupied states are unnecessary to describe
electronic transport. In the context of circuit QED and quantum information
processing, the limit $\Gamma_{N}\ll k_{B}T\ll t_{eh}$ is particularly
relevant since it is desirable that electrons stay a long time in the CPS to
enable their quantum manipulation. In this case, one can calculate the
probability $P_{c}$ of a state $\left\vert c\right\rangle \in\mathcal{E}$ and
the global probability $P_{s}$ of the CPS singly occupied states from the
stationary master equation $(M+M_{rel}+M_{RF})P=0$ with $P=^{t}\{P_{V_{_{1}}%
},P_{V_{_{2}}},P_{T_{+}},P_{T_{-}},P_{T_{0}},P_{s}\}$. The matrix $M$ takes
into account tunnel processes towards the N contacts. Its finite elements are
$M_{sV_{i}}=2v_{i}^{2}\Gamma_{N}$, $M_{V_{i}s}=(1-v_{i}^{2})\Gamma_{N}$,
$M_{sT_{i}}=2\Gamma_{N}$, $M_{V_{i}V_{i}}=-2v_{i}^{2}\Gamma_{N}$,
$M_{T_{i}T_{i}}=-2\Gamma_{N}$ and $M_{ss}=-\Gamma_{N}$, with $v_{i}\in
\lbrack0,1]$ a dimensionless coefficient which depends on $\delta$ (see
Appendix A). The matrix $M_{rel}$ takes into account relaxation processes
between the states of $\mathcal{E}$, due e.g. to phonons. One can use a
rotating wave approximation (RWA) on independent resonances\cite{spectroCPS}
to describe the effect of tone 2 through the matrix $M_{RF}$, with, for
$(c,d)\in\mathcal{E}^{2}$,
\begin{equation}
M_{RF,cd}=\left\vert \varepsilon_{2,cd}\right\vert ^{2}(2\Gamma_{cd}%
/(\omega-\left\vert \omega_{cd}\right\vert )^{2}+\Gamma_{cd}^{2})/\hbar^{2}
\label{tone2}%
\end{equation}
Above, $\Gamma_{cd}$ corresponds to the decoherence rate between the states
$\left\vert c\right\rangle $ and $\left\vert d\right\rangle $. Assuming that
$\Gamma_{cd}$ is limited by relaxation inside $\mathcal{E}$ and tunnel
processes, one can use $\Gamma_{cd}=-(M_{cc}+M_{rel,cc}+M_{dd}+M_{rel,dd})/2$.
In the following, I assume that $\omega_{2}$ is much larger than
$\varepsilon_{B}\Delta_{K\leftrightarrow K^{\prime}}/\Delta_{r}$, and I thus
disregard the elements $M_{RF,T_{0}T_{+[-]}}$. This implies that $\left\vert
T_{0}\right\rangle $ is not populated in the regimes considered below.

Figures 2 to 4 show the variation $\Delta I_{CPS}=I_{CPS}(\varepsilon
_{2})-I_{CPS}(\varepsilon_{2}=0)$ of the CPS input current $I_{CPS}%
=e\Gamma_{N}[2v_{1}^{2},2v_{2}^{2},2,2,2,1].P$ and the variation $\Delta
\omega_{pull}=\omega_{pull}(\varepsilon_{2})-\omega_{pull}(\varepsilon_{2}=0)$
of the cavity frequency pull, versus $\delta$ and $\omega$. Various resonant
lines are visible in $\Delta I_{CPS}$ and $\Delta\omega_{pull}$, for
$\omega_{2}$ equal to $\omega_{V_{1}V_{2}}$, $\omega_{V_{1}T_{\pm}}$, and
$\omega_{V_{2}T_{\pm}}$. Although $\Delta\omega_{cav}$ is dominated by the
charge coupling $\alpha_{V_{_{2}}V_{_{1}}}$ to the cavity, it indirectly
reveals spin-flip transitions $\left\vert V_{1(2)}\right\rangle
\rightleftarrows\left\vert T_{\pm}\right\rangle $ induced by tone 2, due to a
modification of $P_{V_{1(2)}}$. This is similar to the experiment described by
Ref.\cite{Petersson}, where spin transitions in a double quantum dot with a
strong spin-orbit coupling are induced by a classical microwave field applied
locally on the double quantum dot, and read out through the charge coupling to
a coplanar cavity. However, an important difference with Ref.\cite{Petersson}
is that the present work considers a transport situation. This induces
important qualitative modifications of $\Delta\omega_{pull}$, as discussed
below. The presence of an anticrossing due to the resonance $\omega_{2}%
=\omega_{V_{1}V_{2}}$ witnesses the existence of a coherent coupling in the
system. I show below that the characteristics of $\Delta\omega_{pull}$ and
$\Delta I$ point to the coherent injection of split Cooper pairs.

\subsection{Case with no relaxation inside the $\mathcal{E}$ subspace}

One can first neglect relaxation inside $\mathcal{E}$, i.e. $M_{rel,cd}=0$ for
any $c$ and $d$. In this case $\Delta I$ is always positive (Fig.2.a) while
the sign of $\Delta\omega_{pull}$ varies with $\delta$ (Fig.2.b). To
understand this result, one must note that the state of $\mathcal{E}$ which is
the closest to $\left\vert 0,0\right\rangle $ represents a blocking state for
electronic transport, because it has the weakest ability to emit electrons
towards the N contacts. One can check that the blocking state is $\left\vert
V_{_{1}}\right\rangle $ for $\delta<2\Delta_{r}$ and $\left\vert V_{_{2}%
}\right\rangle $ for $\delta>2\Delta_{r}$. This is why $\left\vert V_{_{1}%
}\right\rangle $ [$\left\vert V_{_{2}}\right\rangle $] is the most populated
state for $\delta<2\Delta_{r}$ [$\delta>2\Delta_{r}$] (Fig.2.c). Tone 2 always
give $\Delta I>0$ because it induces transitions towards states which can emit
electrons more easily. The variation $\Delta\omega_{pull}$ behaves differently
because $\omega_{pull}$ is proportional to $P_{V_{1}}-P_{V_{2}}$ (see Eq.
\ref{pull}). I first discuss $\Delta\omega_{pull}$ along the $\left\vert
V_{1}\right\rangle \rightleftarrows\left\vert V_{2}\right\rangle $ resonance.
For $\delta<2\Delta_{r}$, one has $P_{V_{1}}>P_{V_{2}}$ for $\varepsilon
_{2}=0$. Since tone 2 tends to equilibrate $P_{V_{1}}$ and $P_{V_{2}}$ when
$\varepsilon_{2}$ increases (i.e. $P_{V_{1}}-P_{V_{2}}\rightarrow0$), and
since $\mathcal{C}<0$ for the parameters considered in Fig. 2, one obtains
$\Delta\omega_{pull}>0$. Conversely, for $\delta>2\Delta_{r}$, one has
$P_{V_{1}}<P_{V_{2}}$ for $\varepsilon_{2}=0$, thus $\Delta\omega_{pull}<0$.
Hence, $\Delta\omega_{pull}$ changes sign with $\delta$ along the $\left\vert
V_{1}\right\rangle \rightleftarrows\left\vert V_{2}\right\rangle $ resonance,
at $\delta=2\Delta_{r}$. This differs drastically from the usual behavior of a
closed two level system coupled dispersively to a cavity, for which
$\Delta\omega_{pull}$ has a constant sign, because the state with the lowest
energy is always the most populated in the absence of a microwave excitation.
Here, electronic transport provides a way to invert the population of the two
states $\left\vert V_{1}\right\rangle $ and $\left\vert V_{2}\right\rangle $.
This is directly visible in $\Delta\omega_{pull}$ which represents a natural
probe for the population difference $P_{V_{1}}-P_{V_{2}}$. Importantly, the
current signal $\Delta I$ provides a different information, i.e. it indicates
whether tone 2 increases the populations of CPS states with a higher tunnel
rate to the N contacts. Note that both $\Delta\omega_{pull}$ and $\Delta I$
vanish for $\omega_{2}=\omega_{V_{1}V_{2}}$ and $\delta=2\Delta_{r}$ because
the states $\left\vert V_{1}\right\rangle $ and $\left\vert V_{2}\right\rangle
$ play symmetric roles at this point.\begin{figure}[ptb]
\includegraphics[width=1.\linewidth]{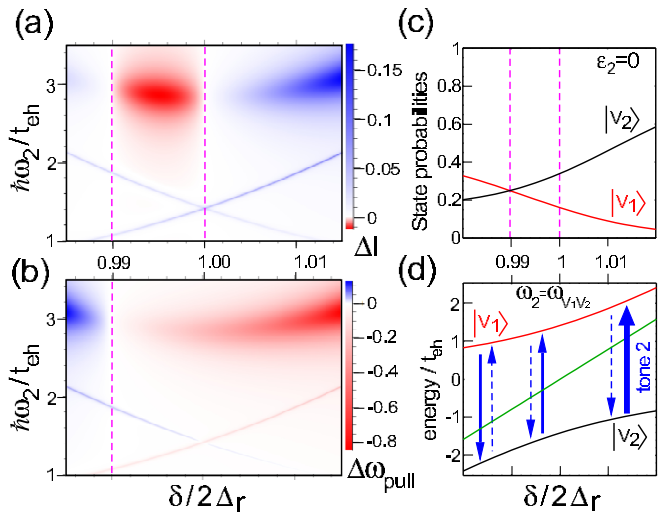}\newline\caption{(a) Current
variation $\Delta I$ versus $\delta$ and $\omega_{2}$ for a strong relaxation
between $\left\vert V_{1}\right\rangle $ and $\left\vert V_{2}\right\rangle $
(b) Corresponding $\Delta\omega_{pull}$ (c) Occupation probabilities of states
$\left\vert V_{1}\right\rangle $ and $\left\vert V_{2}\right\rangle $ for
$\varepsilon_{2}=0$ (d) Scheme illustrating that tone 2 decreases the
population of the lowest state $\left\vert V_{2}\right\rangle $, around
$\delta=2\Delta_{r}$. We have used the same parameters as in Fig.1,
$r=0.55\Gamma_{N}$ and $\varepsilon_{B}=0$.}%
\end{figure}

The resonance $\left\vert V_{1}\right\rangle \rightleftarrows\left\vert
V_{2}\right\rangle $ is broad because the coupling constant $\varepsilon
_{2,V_{1}V_{2}}$ between tone 2 and this transition is large, hence this
transition is saturated, or in other terms, tone 2 leads to $P_{V_{_{1}}%
}=P_{V_{_{2}}}$ for $\omega_{2}=\omega_{V_{1}V_{2}}$. In contrast, the
resonances $\left\vert T_{\pm}\right\rangle \rightleftarrows\left\vert
V_{1(2)}\right\rangle $ appear as thinner lines because they are not saturated
since $\varepsilon_{2,V_{1}T_{\pm}},\varepsilon_{2,T_{\pm}V_{1}}\ll
\varepsilon_{2,V_{1}V_{2}}$ (see Fig.2). Spin-orbit coupling enables tone 2 to
populate the states $\left\vert T_{\pm}\right\rangle $ which are unoccupied
for $\varepsilon_{2}=0$. This is why one keeps $\Delta\omega_{pull}>0$ along
the $\left\vert V_{1}\right\rangle \rightleftarrows\left\vert T_{\pm
}\right\rangle $ resonance and $\Delta\omega_{pull}<0$ along the $\left\vert
V_{2}\right\rangle \rightleftarrows\left\vert T_{\pm}\right\rangle $
resonance, for any value of $\delta$. Furthermore, $\Delta I_{CPS}$ remains
positive along both resonances, because the triplet states have no $\left\vert
0,0\right\rangle $ component, and they thus emit electrons to the N reservoirs
faster than $\left\vert V_{1(2)}\right\rangle $, for any $\delta$.

When one applies a DC magnetic field $B$ to the circuit, the resonant lines
involving the triplet states split into two lines due to Eq. (\ref{Et}), while
the $\left\vert V_{1}\right\rangle \rightleftarrows\left\vert V_{2}%
\right\rangle $ resonances are unchanged because they involve the singlet
state (Fig.3). Using $B\neq0$ can thus be instrumental to reveal the spin
structure of the system in an experiment, and confirm that the anticrossing
given by the $\left\vert V_{1}\right\rangle \rightleftarrows\left\vert
V_{2}\right\rangle $ resonance is due to the injection of spin singlet Cooper pairs.

\subsection{Effect of relaxation between the states $\left\vert V_{1}%
\right\rangle $ and $\left\vert V_{2}\right\rangle $}

In practice, relaxation and dephasing can occur between the different CPS
states. Dephasing should only modify the visibility of the resonant lines,
through Eq.(\ref{tone2}). In contrast, relaxation could induce qualitative
modifications of $\Delta I$ and $\Delta\omega_{pull}$. For simplicity, in the
following, I use $M_{rel,V_{2}V_{1}}=r$ and the other elements of $M_{rel}$
equal to $0$, because $\left\vert V_{1}\right\rangle $ and $\left\vert
V_{2}\right\rangle $ have the same spin symmetry, thus the transition
$\left\vert V_{1}\right\rangle \rightleftarrows\left\vert V_{2}\right\rangle $
should be affected by relaxation induced e.g. by phonons. Figure 4 shows
$\Delta I$ and $\Delta\omega_{pull}$ for the same parameters as in Fig.2, and
$r$ finite. Around $\delta=2\Delta_{r}$, $\left\vert V_{2}\right\rangle $ is
the most populated state. Hence, $\Delta I$ now changes sign along the
$\left\vert V_{1}\right\rangle \rightleftarrows\left\vert V_{2}\right\rangle $
resonance at the point $\delta=2\Delta_{r}$, while $\Delta\omega_{pull}$
remains negative. If $r<2\Gamma_{N}$, a sign change of $\Delta\omega_{pull}$
persists (see Fig.4.b) for a value of $\delta$ smaller than $2\Delta_{r}$,
where $\left.  P_{V_{1}}\right\vert _{\varepsilon_{2}=0}=\left.  P_{V_{2}%
}\right\vert _{\varepsilon_{2}=0}$ (see Fig.4.c). This effect goes together
with a second sign change of $\Delta I$ (Fig.4.a). If $r>2\Gamma_{N}$,
$\Delta\omega_{pull}$ keeps a constant sign along the whole $\left\vert
V_{1}\right\rangle \rightleftarrows\left\vert V_{2}\right\rangle $ resonance
(not shown). However, even for $r\gg\Gamma_{N}$, $\Delta\omega_{pull}$ shows a
strong asymmetry with respect to $\delta=2\Delta_{r}$, similar to what shown
in Fig.4.b for $\delta/2\Delta_{r}>0.99$, because $\left\vert V_{2}%
\right\rangle $ is the blocking state for $\delta>2\Delta_{r}$ only. Hence,
even in the presence of internal relaxation in the CPS, $\Delta\omega_{pull}$
shows a behavior which is very specific to a transport situation.

\subsection{Expected amplitude of the signals}

It is important to point out that the above effects are already within
experimental reach. Joint measurements of the current through a nanocircuit
and the corresponding cavity frequency pull are now realized commonly in
experiments combining nanocircuits and coplanar microwave
cavities\cite{Delbecq,DQD,Viennot}. For the realistic parameters used in Figs.
2, 3 and 4 (see Refs.\cite{CPSexp,Jespersen,Liang,Kuemmeth}), the magnitude of
$\Delta\omega_{pull}[\Delta I]$ is set by the scales $\omega_{0}\sim2\pi
\times40kHz$ [$e\Gamma_{N}\sim20pA$]. Hence, these signals are accessible
experimentally with present techniques\cite{Viennot}. The above model and
parameters are compatible with cavity quality factors $Q\sim1000$ obtained
presently in Hybrid Circuit QED. This works considers CNT-based devices which
are the most advanced systems for Cooper pair splitting\cite{Schindele}, but
similar results are expected with other types of nanoconductors.

\section{Conclusion}

To summarize, Hybrid Circuit QED provides a direct access to the coherence of
Cooper pair injection in the CPS. This coherence is revealed by an
anticrossing in the cavity frequency pull, which can be discriminated from all
other possible anticrossings because of various unusual specificities. First,
this anticrossing is visible along the $\delta$ axis, which necessarily points
to processes involving electron pairs split between the two dots. Second, it
displays sign changes or asymmetries with $\delta$, which reveal a population
inversion due to out-of equilibrium transport. These properties are difficult
to mimic without an exchange of particles with a superconducting reservoir.
Third, the splitting of the cavity frequency pull with a magnetic field
reveals the spin structure of the two-particle states involved. Note that
these results do not represent a direct proof for the conservation of spin
entanglement in the CPS, but it seems unlikely to have spin entanglement
conservation without coherent pair injection. Observing the coherent pair
injection through the cavity frequency pull can thus be an instrumental step
towards the realization of a fully coherent CPS. More generally, this work
illustrates that Hybrid Circuit QED provides a rich tool to study electronic
transport in nanostructures.

Note that the present work considers a limit where one can disregard single
quasiparticle transport from the superconducting contact to the dots, as well
as Cooper pair injection in a single dot or other parasitic
processes\cite{Sauret}. In a real experiment, these processes could become
significant depending on the device parameters. Nevertheless, this should
modify only quantitatively the properties of the anticrossing induced by
$t_{eh}$, if one achieves a sufficient Cooper pair splitting rate.

\textit{I acknowledge fruitful discussions with T. Kontos. This work has been
financed by the EU FP7 project SE2ND[271554] and the ANR-NanoQuartet
[ANR12BS1000701] (France).}

\section{Appendix A: Expression of the CPS eigenstates}

I denote with $\left\vert \tau\sigma,\tau^{\prime}\sigma^{\prime}\right\rangle
$ the CPS state with one electron with spin $\sigma$ in orbital $\tau$ of dot
$L$ and one electron with spin $\sigma^{\prime}$ in orbital $\tau^{\prime}$ of
dot $R$. By definition, the spin states $\sigma\in\{\uparrow,\downarrow\}$ are
along the carbon nanotube axis, and parrallel to the effective field $\vec
{h}_{so}$ produced by the spin orbit coupling (term in $\Delta_{so}$). The
five eigenstates of the subspace $\mathcal{E}$ discussed in the main text
are:
\begin{gather}
\left\vert V_{1}\right\rangle =\sqrt{1-v_{1}^{2}}\left\vert 0,0\right\rangle
+v_{1}\left\vert \mathcal{S}\right\rangle \\
\left\vert V_{2}\right\rangle =\sqrt{1-v_{2}^{2}}\left\vert 0,0\right\rangle
+v_{2}\left\vert \mathcal{S}\right\rangle \\
\left\vert T_{-1}\right\rangle =(\left\vert \mathcal{T}_{0}\right\rangle
-\left\vert \mathcal{T}_{-}\right\rangle )/\sqrt{2}\\
\left\vert T_{+1}\right\rangle =(\left\vert \mathcal{T}_{0}\right\rangle
+\left\vert \mathcal{T}_{-}\right\rangle )/\sqrt{2}\\
\left\vert T_{0}\right\rangle =\left\vert \mathcal{T}_{+}\right\rangle
\end{gather}
Above, the state $\left\vert \mathcal{T}_{0}\right\rangle $ correspond to a
generalized triplet state with zero spin along the nanotube axis, and
$\left\vert \mathcal{T}_{+}\right\rangle $ and $\left\vert \mathcal{T}%
_{-}\right\rangle $ correspond to coherent superpositions of triplet states
with equal spins, i.e.
\begin{align}
\left\vert \mathcal{S}\right\rangle  &  =\sum_{\sigma}\left\{  \frac{1}%
{2}(\frac{\Delta_{so}}{\Delta_{r}}-\sigma)\left\vert \mathcal{C}_{-}%
(K\sigma,K^{\prime}\bar{\sigma})\right\rangle \right\}  \\
&  +\frac{\Delta_{K/K^{\prime}}}{2\Delta_{r}}%
%TCIMACRO{\dsum \limits_{\tau}}%
%BeginExpansion
{\displaystyle\sum\limits_{\tau}}
%EndExpansion
\left\vert \mathcal{C}_{-}(\tau\uparrow,\tau\downarrow)\right\rangle
\end{align}%
\begin{align}
\left\vert \mathcal{T}_{0}\right\rangle  &  =\sum_{\sigma}\frac{1}{2}%
(\sigma\frac{\Delta_{so}}{\Delta_{r}}-1)\left\vert \mathcal{C}_{+}%
(K\sigma,K^{\prime}\bar{\sigma})\right\rangle \label{T0}\\
&  +\frac{\Delta_{K/K^{\prime}}}{2\Delta_{r}}%
%TCIMACRO{\dsum \limits_{\tau}}%
%BeginExpansion
{\displaystyle\sum\limits_{\tau}}
%EndExpansion
\left\vert \mathcal{C}_{+}(\tau\uparrow,\tau\downarrow)\right\rangle
\end{align}%
\begin{align}
\left\vert \mathcal{T}_{+}\right\rangle  &  =\sum_{\sigma}\frac{1}{2}\left(
\frac{\Delta_{so}}{\Delta_{r}}-\sigma\right)  \frac{\left\vert K\sigma
,K\sigma\right\rangle -\left\vert K^{\prime}\bar{\sigma},K^{\prime}\bar
{\sigma}\right\rangle }{\sqrt{2}}\label{Tplus}\\
&  +\sum_{\sigma}\sigma\frac{\Delta_{K/K^{\prime}}}{2\Delta_{r}}\left\vert
\mathcal{C}_{+}(K\sigma,K^{\prime}\sigma)\right\rangle
\end{align}%
\begin{align}
\left\vert \mathcal{T}_{-}\right\rangle  &  =\sum_{\sigma}\frac{1}{2}\left(
1-\frac{\Delta_{so}}{\Delta_{r}}\sigma\right)  \frac{\left\vert K\sigma
,K\sigma\right\rangle +\left\vert K^{\prime}\bar{\sigma},K^{\prime}\bar
{\sigma}\right\rangle }{\sqrt{2}}\label{Tmoins}\\
&  -\sum_{\sigma}\frac{\Delta_{K/K^{\prime}}}{2\Delta_{r}}\left\vert
\mathcal{C}_{+}(K\sigma,K^{\prime}\sigma)\right\rangle
\end{align}
and%
\begin{equation}
v_{1(2)}=\frac{2t_{eh}}{\sqrt{8t_{eh}^{2}+d(d\mp\sqrt{8t_{eh}^{2}+d^{2}})}%
}\label{v12}%
\end{equation}
with $d=\delta-2\Delta_{r}$. I have used above $\left\vert \mathcal{C}_{\pm
}(\tau\sigma,\tau^{\prime}\sigma^{\prime})\right\rangle =(\left\vert
\tau\sigma,\tau^{\prime}\sigma^{\prime}\right\rangle \pm\left\vert
\tau^{\prime}\sigma^{\prime},\tau\sigma\right\rangle )/\sqrt{2}$. Note that
$\sigma=\pm1$ stands for spin states $\sigma\in\{\uparrow,\downarrow\}$ in
algebraic expressions, with $\bar{\sigma}=-\sigma$. The eigenstates
$\left\vert T_{-1}\right\rangle $ and $\left\vert T_{+1}\right\rangle $ of the
full system correspond to a superposition of $\left\vert \mathcal{T}%
_{0}\right\rangle $ and $\left\vert \mathcal{T}_{-}\right\rangle $, due to the
presence of the magnetic field $\vec{B}$ which is perpendicular to $\vec
{h}_{so}$.

\section{Appendix B: Approximations}

This section discusses various approximations used in the main text.

\subsection{Photon-induced transition between CPS singly occupied states}

Photon-induced transition inside the CPS singly occupied charge sector could
modify $\Delta\omega_{cav}$ and the reaction of the CPS to tone 2, in
principle. To discuss this possibility it is useful to recall that one has
typically $t_{ee}$, $\Delta_{r}\gg t_{eh}$, $\hbar\omega_{cav}$. I furthermore
assume that $\varepsilon_{B}\ll\hbar\omega_{cav},\hbar\omega_{2}$. One can
check that photon-induced transitions inside the CPS singly occupied charge
sector correspond to frequencies of the order of $2\Delta_{r}$, $2t_{ee}$, or
$\varepsilon_{B}$. The two first values are typically too large and the last
one too small to enable an excitation inside the singly occupied charge sector
by tone 2, because of the limited frequency range of microwave sources
($\hbar\omega_{2}\ll t_{ee}$, $\Delta_{r}$) and because I assume $\hbar
\omega_{2}\gg\varepsilon_{B}$. Regarding $\Delta\omega_{pull}$, one can expect
a significant contribution from the charge couplings $\lambda_{L(R)}$ only.
One can check that photon-induced transitions caused by $\lambda_{L(R)}$
inside the singly occupied charge sector have frequencies $2t_{ee}$ which is
typically huge compared to $t_{eh}$ and $\hbar\omega_{cav}$. Therefore the
contribution of these transitions to $\Delta\omega_{pull}$ can be disregarded
in comparison with the contribution (\ref{pull}) from the main text.

\subsection{RWA on independent resonances}

The RWA on independent resonances requires that the various resonances induced
by tone 2 are sufficiently separated. This is not justified at the crossing
between the different thin resonances in Figs.2 to 4. Nevertheless,
corrections are expected in a very small fraction of the parameters space,
barely visible in Figs. 2 to 4. The related physics goes beyond the scope of
this article.

\subsection{CPS/cavity coupling elements}

For simplicity, Eq.(\ref{coupl}) of the main text restricts the symmetry of
the spin-flip and orbit-change terms of $h_{c}$. There can be extra terms with
other symmetries, depending on the microscopic details of the carbon nanotube
quantum dots. The terms used in the main text lead to the most interesting
effects expected in the CPS/cavity system. For the spin-flip terms in $h_{c}$,
extra contributions in $\mathbf{i}\sigma\tau\tilde{\lambda}_{i}d_{i\tau\sigma
}^{\dag}d_{i\tau\overline{\sigma}}$ or $\tau\tilde{\lambda}_{i}d_{i\tau\sigma
}^{\dag}d_{i\tau\overline{\sigma}}$ are compatible with the hermicity of
$H_{tot}$, but this does not modify the coupling between the states of
$\mathcal{E}$. An extra contribution in $\tilde{\lambda}_{i}d_{i\tau\sigma
}^{\dag}d_{i\tau\overline{\sigma}}$ would add couplings $\alpha_{T_{0}%
V_{1(2)}}=v_{1(2)}\mathbf{i}(\tilde{\lambda}_{L}-\tilde{\lambda}_{R}%
)\Delta_{K\leftrightarrow K^{\prime}}/\Delta_{r}$ between the states
$\left\vert V_{1(2)}\right\rangle $ and $\left\vert T_{0}\right\rangle $. This
could produce extra thin resonant lines in Figs. 3.a and 3.b for $\omega
_{2}=\omega_{V_{1}T_{0}}$ and $\omega_{2}=\omega_{T_{0}V_{2}}$. This effect
can be included straightforwardly in the system description.

For the photon-induced orbit-changes, an imaginary contribution to $h_{c}$
with the form $\mathbf{i}\tau\tilde{\alpha}_{i}d_{i\tau\sigma}^{\dag}%
d_{i\bar{\tau}\sigma}$ or $\mathbf{i}\tau\tilde{\alpha}_{i}d_{i\tau\sigma
}^{\dag}d_{i\bar{\tau}\sigma}$ is possible, in principle, but this does not
modify the coupling between the states of $\mathcal{E}$. A contribution with
the form $\tilde{\alpha}_{i}\sigma d_{i\tau\sigma}^{\dag}d_{i\bar{\tau}\sigma
}$ would lead to a renormalisation of Eqs.(\ref{alpha1}) and (\ref{alpha1b}),
i.e. one should replace $\mp\mathbf{i}(\lambda_{L}-\lambda_{R})$ by
$\mp\mathbf{i}(\lambda_{L}-\lambda_{R})\mathbf{+}(\tilde{\alpha}_{L}%
-\tilde{\alpha}_{R})$ and $\mathbf{i}(\lambda_{L}+\lambda_{R})$ by
$\mathbf{i}(\lambda_{L}+\lambda_{R})\mp(\tilde{\alpha}_{L}+\tilde{\alpha}%
_{R})$. This would affect only quantitatively the results presented in this paper.

In any case, the CPS/cavity charge couplings $\beta_{L(R)}$ are expected to be
dominant, so that the spin-flip and orbit-change couplings will not affect the
$\left\vert V_{1}\right\rangle \rightleftarrows\left\vert V_{2}\right\rangle $
resonance, but rather control the thin resonant lines in Figs. 2 to 4.


\begin{thebibliography}{99}                                                                                               %


\bibitem {Recher:01}P. Recher, E. V. Sukhorukov, and D. Loss, Phys. Rev. B
\textbf{63}, 165314 (2001).

\bibitem {CPSexp}L. Hofstetter, S. Csonka, J. Nyg\aa rd and C.
Sch\"{o}nenberger, Nature \textbf{461}, 960 (2009); L. G. Herrmann, F.
Portier, P. Roche, A. Levy Yeyati, T. Kontos, and C. Strunk, Phys. Rev. Lett.
\textbf{104}, 026801 (2010); L. Hofstetter, S. Csonka, A. Baumgartner, G.
F\"{u}l\"{o}p, S. d'Hollosy, J. Nyg\aa rd, and C. Sch\"{o}nenberger, Phys.
Rev. Lett. \textbf{107}, 136801 (2011); J. Schindele, A. Baumgartner, and C.
Sch\"{o}nenberger, Phys. Rev. Lett. \textbf{109}, 157002 (2012); L. G.
Herrmann, P. Burset, W. J. Herrera, F. Portier, P. Roche, C. Strunk, A. Levy
Yeyati, and T. Kontos, arXiv:1205.1972; A. Das, Y. Ronen, M. Heiblum, D.
Mahalu, A. V. Kretinin and H. Shtrikman, Nature Comm. 3, Article number: 1165 (2012).

\bibitem {Schindele}J. Schindele, A. Baumgartner, R. Maurand, M. Weiss, and C.
Sch\"{o}nenberger, Phys. Rev. B, \textbf{89}, 045422 (2014).

\bibitem {NewTools}T. Martin, Phys. Lett. A \textbf{220}, 137 (1996); M. P.
Anantram and S. Datta, Phys. Rev. B \textbf{53}, 16390 (1996); G. Burkard, D.
Loss, and E. V. Sukhorukov, Phys. Rev. B \textbf{61}, R16303 (2000); S.
Kawabata, J. Phys. Soc. Jpn. \textbf{70}, 1210 (2001); G. B. Lesovik, T.
Martin, and G. Blatter, Eur. Phys. J. B \textbf{24}, 287 (2001).

\bibitem {Schroer}A. Schroer, B. Braunecker, A. Levy Yeyati, and P. Recher, arXiv:1404.4524.

\bibitem {Sauret}O. Sauret, D. Feinberg, and T. Martin, Phys. Rev. B 70,
245313 (2004).

\bibitem {Eldridge}J. Eldridge, M. G. Pala, M. Governale, and J. K\"{o}nig,
Phys. Rev. B 82, 184507 (2010).

\bibitem {QED}H. Mabuchi and A. Doherty, Science \textbf{298}, 1372 (2002).

\bibitem {Raimond}J.M. Raimond, M. Brune and S. Haroche, Rev. Mod.
Phys.\textbf{\ 73}, 565 (2001).

\bibitem {Wallraff}A. Wallraff, D. I. Schuster, A. Blais, L. Frunzio, R.- S.
Huang, J. Majer, S. Kumar, S. M. Girvin \& R. J. Schoelkopf, Nature
\textbf{431}, 162 (2004).

\bibitem {Delbecq}M. R. Delbecq, V. Schmitt, F. D. Parmentier, N. Roch, J. J.
Viennot, G. F\`{e}ve, B. Huard, C. Mora, A. Cottet, and T. Kontos, Phys. Rev.
Lett. \textbf{107}, 256804 (2011); M.R. Delbecq, L.E. Bruhat, J.J. Viennot, S.
Datta, A. Cottet and T. Kontos, Nature Communications \textbf{4}, Article
number: 1400 (2013).

\bibitem {DQD}T. Frey, P. J. Leek, M. Beck, A. Blais, T. Ihn, K. Ensslin, and
A. Wallraff, Phys. Rev. Lett. \textbf{108}, 046807 (2012); M. D. Schroer, M.
Jung, K. D. Petersson, and J. R. Petta, Phys. Rev. Lett. \textbf{109}, 166804
(2012); H. Toida, T. Nakajima, and S. Komiyama, Phys. Rev. Lett. \textbf{110},
066802 (2013); J. Basset, D.-D. Jarausch, A. Stockklauser, T. Frey, C. Reichl,
W. Wegscheider, T. M. Ihn, K. Ensslin, and A. Wallraff, Phys. Rev. B
\textbf{88}, 125312 (2013); G.-W. Deng, D. Wei, J.R. Johansson, M.-L. Zhang,
S.-X. Li, H.-O. Li, G. Cao, M. Xiao, T. Tu, G.-C. Guo, H.-W. Jiang, F. Nori,
G.-P. Guo, arXiv:1310.6118.

\bibitem {Petersson}K. D. Petersson, L. W. McFaul, M. D. Schroer, M. Jung, J.
M. Taylor, A. A. Houck and J. R. Petta, Nature \textbf{490}, 380 (2012).

\bibitem {Viennot}J. J. Viennot, M. R. Delbecq, M. C. Dartiailh, A. Cottet,
and T. Kontos, Phys. Rev. B \textbf{89}, 165404 (2014).

\bibitem {CottetSST}A. Cottet, T. Kontos, S. Sahoo, H. T. Man, M.-S. Choi, W.
Belzig, C. Bruder, A. F. Morpurgo and C. Sch\"{o}nenberger, Semicond. Sci.
Technol. \textbf{21}, S78 (2006).

\bibitem {Franceschi}S. De Franceschi, L. Kouwenhoven, C. Sch\"{o}nenberger
and W. Wernsdorfer, Nature Nanotechnology \textbf{5}, 703 (2010).

\bibitem {Qubits}M. Trif, V. N. Golovach, and D. Loss, Phys. Rev. B
\textbf{77}, 045434 (2008); A. Cottet and T. Kontos, Phys. Rev. Lett.
\textbf{105}, 160502 (2010); P.-Q. Jin, M. Marthaler, A. Shnirman, and G.
Sch\"{o}n, Phys. Rev. Lett. \textbf{108}, 190506 (2012); X. Hu, Y.-X. Liu, and
F. Nori, Phys. Rev. B \textbf{86}, 035314 (2012); C. Kloeffel, M. Trif,
P.Stano, and D. Loss, Phys. Rev. B \textbf{88}, 241405(R) (2013).

\bibitem {CKLY}A. Cottet, T. Kontos, and A. Levy Yeyati, Phys. Rev. Lett.
\textbf{108}, 166803 (2012).

\bibitem {MajoDirect}M. Trif and Y. Tserkovnyak, Phys. Rev. Lett.
\textbf{109}, 257002 (2012); A. Cottet, T. Kontos and B. Dou\c{c}ot, Phys.
Rev. B \textbf{88}, 195415 (2013); T. L. Schmidt, A. Nunnenkamp, and C.
Bruder, Phys. Rev. Lett. \textbf{110}, 107006 (2013).

\bibitem {MajoTransmon}F. Hassler, A. R. Akhmerov and C. W. J Beenakker, New
J. Phys. \textbf{13}, 095004 (2011); T. Hyart, B. van Heck, I. C. Fulga, M.
Burrello, A. R. Akhmerov, and C. W. J. Beenakker, Phys. Rev. B \textbf{88},
035121 (2013); C. M\"{u}ller, J. Bourassa and A. Blais, Phys. Rev. B
\textbf{88}, 235401 (2013); E. Ginossar and E.Grosfeld Nat. Commun. 5, 4772 (2014).

\bibitem {Lasing}L. Childress, A. S. S\o rensen, and M. D. Lukin, Phys. Rev. A
\textbf{69}, 042302 (2004); P.-Q. Jin, M. Marthaler, J. H. Cole, A. Shnirman,
and G. Sch\"{o}n, Phys. Rev. B \textbf{84}, 035322 (2011); M. Kulkarni, O.
Cotlet, and H. E. T\"{u}reci, arXiv:1403.3075.

\bibitem {2DQDS}C. Bergenfeldt and P. Samuelsson, Phys. Rev. B \textbf{87},
195427 (2013); N. Lambert, C. Flindt, and Franco Nori, Europhys. Lett.
\textbf{103}, 17005 (2013); L D Contreras-Pulido, C Emary, T Brandes and
Ram\'{o}n Aguado, New Journal of Physics \textbf{15}, 095008 (2013); C.
Bergenfeldt, P. Samuelsson, B. Sothmann, C. Flindt, and M. B\"{u}ttiker, Phys.
Rev. Lett. \textbf{112}, 076803, (2014).

\bibitem {Liu}Y.-Y. Liu, K.\thinspace D. Petersson, J. Stehlik, J.\thinspace
M. Taylor, and J.\thinspace R. Petta, Phys. Rev. Lett. \textbf{113}, 036801 (2014).

\bibitem {Jespersen}T. S. Jespersen, K. Grove-Rasmussen, J. Paaske, K. Muraki,
T. Fujisawa, J. Nyg\aa rd and K. Flensberg, Nature Physics \textbf{7}, 348 (2011).

\bibitem {Liang}W. Liang, M. Bockrath and H. Park, Phys. Rev. Lett.
\textbf{88}, 126801 (2002).

\bibitem {Kuemmeth}F. Kuemmeth, S. Ilani, D. C. Ralph and P. L. McEuen Nature
\textbf{452}, 448 (2008).

\bibitem {Palyi}A. P\'{a}lyi and G. Burkard, Phys. Rev. B \textbf{82}, 155424
(2010); A. P\'{a}lyi and G. Burkard, Phys. Rev. Lett. \textbf{106}, 086801 (2011).

\bibitem {CKD}A. Cottet et al., to be published elsewhere.

\bibitem {spectroCPS}A. Cottet, Phys. Rev. B \textbf{86}, 075107 (2012).
\end{thebibliography}
\end{document}